\def \K{~\rm{K}}

\def \yr{~\rm{yr}}

\documentclass[12pt,preprint]{aastex}
\begin{document}

\title{DEFINING THE TERMINATION OF THE ASYMPTOTIC GIANT BRANCH}

\author{Noam Soker\altaffilmark{1}}

\altaffiltext{1}{Dept. of Physics, Technion, Haifa 32000, Israel;
soker@physics.technion.ac.il.}

\begin{abstract}
I suggest a theoretical quantitative definition for the termination of the
asymptotic giant branch (AGB) phase and the beginning of the post-AGB phase.
I suggest that the transition will be taken to occur when the ratio of the dynamical
time scale to the the envelope thermal time scale,
$Q \equiv {\tau_{\rm dyn}}/{\tau_{\rm KH-env}}$, reaches its maximum value.
Time average values are used for the different quantities, as the criterion
does not refer to the short time-scale variations occurring on the
AGB and post-AGB, e.g., thermal pulses (helium shell flashes) and magnetic activity.
Along the entire AGB the value of $Q$ increases, even when the star starts to
contract. Only when a rapid contraction starts does the value of $Q$ start to decrease.
This criterion captures the essence of the transition from the AGB to the post AGB phase,
because $Q$ is connected to the stellar effective temperature, reaching its maximum
value at $T \simeq 4000-6000 \K$, it is related to the mass loss properties,
and it reaches its maximum value when rapid contraction starts and envelope mass is very low.
\end{abstract}

\section{INTRODUCTION}
\label{sec:intro}

The transition of asymptotic giant branch (AGB) stars to the early planetary nebula (PN)
stage is a poorly understood phase of stellar evolution.
It is agreed upon that a rapid drop in mass loss rate accompanies the transition from the
AGB to the post-AGB phase.
However, no well defined objective criterion exists for the transition from the
AGB to the post-AGB phase (the termination of the AGB).
Bl\"ocker (1995), for example, used both mass loss rate and the stellar pulsation period,
i.e., a dynamical property of the star.
He took the mass loss rate to decrease with decreasing pulsating period, and defined
the zero-age post-AGB phase when the pulsation period is 50 days.
Below I will suggest a new property to theoretically define the transition from
the AGB to the post-AGB phase.
It is based on both the thermal and dynamical properties of the star, and
is connected to the mass loss rate and mass loss inhomogeneities.

Observationally, AGB stars are well defined,  and {\it visible } well developed
post-AGB stars can be clearly defined in principle, but practically there are
many difficulties (e.g., Hrivnak et al. 1989; Szczerba et al. 2007; Suarez et al. 2006).
Still, there is no theoretical definition of the transition.
Recently, Szczerba et al. (2007) classified many post-AGB stars.
However, they classified mainly unobscured stars in the IR and/or the visible band, most likely
well after they have left the AGB.
Observationally, most stars are likely to be obscured during the transition because
of a high mass loss rate.
Several parameters were proposed to define the transition from the AGB to the post-AGB
phase, but each one of them has some problems.
\\
{\it A drop in the mass loss rate.} This criterion (e.g., Suarez et al. 2006) captures the
essence of the transition.
However, what mass loss rate should be used to mark the transition?
What if the star rotates and/or interacts with a binary companion such that
mass loss rate depends on the companion?
\\
{\it Optical depth.} Suggestions have been made that the transition occurs
  when the totally obscured AGB star becomes visible again. But at what wavelength?
  Another problem with this definition is that it depends on the viewing angle if
  the mass loss geometry is not spherical.
In addition, low mass AGB stars with low metallicity might never become totally obscured.
\\
{\it Pulsation period.} Although mass loss rate is tightly coupled to pulsation,
it is not clear what pulsation period should be used. Also, AGB stars with different
total masses and luminosity can have
the same pulsation period at very different effective temperatures.
The usage of pulsation  (e.g., Bl\"ocker 1995) includes in it the dynamical time of the star,
but no `natural' transition value exists. In the presently proposed criterion the
dynamical time is included with a quantitative measure.
\\
{\it A change in the dominating mass loss mechanism. } On the AGB the  mass loss process is
     that of pulsations coupled with radiation pressure on dust, while for the central stars
    of PNs it is mainly radiation pressure on ions.  The idea is that the transition
    is defined when the dominate mass loss process switches from pulsation and radiation
    pressure on dust to radiation pressure on ions.
    There are two main problems with this. Firstly, the physics is not well understood
    to connect this transition to stellar evolutionary codes. Secondly, interaction with
    a companion can be the dominate mass loss mechanism in many post-AGB
    stars, either via tidal interaction or a common envelope.
\\
 {\it Effective temperature.} It is not clear what temperature to use. There is no
`natural' temperature for any physical effect, although the transition occurs
around an effective temperature of $T \simeq 5000 \K$ (e.g. Sch\"onberner 1981).
Even dust formation can cease at different temperatures, depending on the metallicity
of the envelope.
Vassiliadis \& Wood (1994) took the transition to occur when the effective temperature is
twice the minimum temperature the star can reach on the AGB, but no physical reason is given for that.
\\
{\it Contraction. } Contraction cannot be used because the star starts to contract
 before it leaves the AGB.
\\
{\it Rapid contraction. }
The criterion of a rapid contraction, with time or with decreasing envelope mass,
captures the essence of the transition, but a quantitative value is not easy to define.
One can use the logarithmic derivative of the stellar radius with envelope mass
\begin{equation}
\delta \equiv \frac{d \ln R }{d \ln {M_{\rm env}} } .
\label{delta1}
\end{equation}
But $\delta$ changes monotonically in the relevant temperature (radius) range,
and it is not clear what value should be used, although $\delta=1$ might be a natural
choice (Frankowski, A. 2007, private communication).
Alternatively, one can define the transition to occur when the magnitude of
the second logarithmic derivative of the stellar radius with envelope mass
\begin{equation}
C \equiv \vert \frac{d^2 \ln R }{d \ln {M_{\rm env}}^2 } \vert
\label{C1}
\end{equation}
reaches its maximum value.
This occurs when the contraction changes from the AGB type behavior to the post-AGB one.
Examining some models show that this occurs at an effective temperature of $\sim 7000-9000 \K$,
which is hotter than the usually assumed transition point (Sch\"onberner \& Bl\"ocker 1993),
e.g.,
Szczerba et al. (2001) and Tylenda et al. (2001) who listed G and K stars as post AGB stars.
The criterion proposed in the next section includes in it the beginning of the rapid contraction
with envelope mass, and does it with a quantitative physical definition.
In any case, it seems that the criterion of maximum $C$ is similar in some aspects
to the criterion suggested in the next section (but not the quantitative value of transition).
\\
{\it Topology in the $U-V$ plane.}
One defines the quantities $V\equiv 4 \pi r^3 \rho/M_r$ and $U \equiv G M_r \rho /rP$,
where $M_r$ is the mass inner to radius $r$ in the star, and the other symbols have their
usual meaning.
One can draw the structure of the star in the $U-V$ plane. As discussed in
detailed by Sugimoto \& Fujimoto (2000), a structural curve with a loop corresponds
to a giant-like structure, while that without a loop corresponds to a dwarf-like structure.
Equivalently, one can consider the variation of $W \equiv V/W$ within the star.
If it changes (inside the star) monotonically then there is no loop.
We can try to apply this criterion to the post AGB star. For a very low envelope mass
$M_r$ is constant in the envelope, and $W=GM_r^2/4 \pi r^4 P \sim (r^4P)^{-1}$.
I examine the structure of the model with the envelope mass of $5.74 \times 10^{-4}$
from Soker (1992), that has an effective temperature of $7200 K$.
The envelope pressure profile changes from $P \sim r^{-5}$ in the range
$r \la 3 R_\odot$ to $P \sim r^{-3.7}$ in the range $4 \la r \la 20 R_\odot$.
Namely, $W$ increases with $r$ for $r \la 3 R_\odot$, and decreases with increasing $r$
in the range $4 \la r \la 20 R_\odot$. A loop in the $U-V$ plane does exist for this model.
Therefore, the loop in the $U-V$ plane disappears too late in the post AGB evolution.
An interesting property is that $V$ and $U$ are related to dynamical and thermal
properties of the envelope, however, it is not straightforward to connect
them to relevant properties of the post-AGB envelope.
\\
{\it Disappearance of the envelope convective zone.}
Although the convective zone becomes thinner and thinner as the star shrinks,
it disappear only when the star is very hot (e.g., Soker 1992).
\\
{\it End of shell burning.} This criterion has not been suggested, but it is listed here for the
sake of clarity. It is not a good criterion as nuclear burning ceases when the star is
hot (e.g. Harpaz \& Kovetz 1981; Kovetz \& Harpaz 1981;  Sch\"onberner 1981).

\section{THE PROPOSED CRITERION}
\label{sec:criterion}

When the envelope mass is low, the Kelvin-Helmholtz time of the envelope is given by
(Sch\"onberner \& Bl\"ocker 1993)
\begin{equation}
\tau_{\rm KH-env} = \frac{G M_c}{L} \int \frac{ 4 \pi r^2\rho(r)}{r} {dr},
\label{tkh1}
\end{equation}
where $M_c$ is the core mass, $L$ is the stellar luminosity, $R$ is the stellar radius,
and $\rho$ is the density in the envelope.
On the upper AGB the density profile in most of the envelope (beside regions
very close to the core that contain very little mass)
can be approximated by $\rho \propto r^{-2}$ (e.g., Soker 1992).
For the inner radius of the envelope, and in particular for the convective part,
we can take $r_0 \sim 1 R_\odot$, and so $\beta_s=\ln (R/r_0) \simeq 6$ in the integration of
equation (\ref{tkh1}).
I defined a parameter $\beta_s$ that depends on the exact structure of the envelope.
For the response of the envelope alone, the internal energy of the envelope should also
be considered. This will reduce the required time to supply energy or to remove energy,
and will reduce somewhat the effective value of $\beta_s$.
I will therefore scale it with $\beta_s=5$.
Scaling the different variables of upper AGB stars gives
\begin{equation}
\tau_{\rm KH-env} \simeq 6
\left( \frac{\beta_s}{5} \right)
\left( \frac{M_c}{0.6 M_\odot} \right)
\left( \frac{M_{\rm env}}{0.1 M_\odot} \right)
\left( \frac{L}{5000 L_\odot} \right)^{-1}
\left( \frac{R}{300 R_\odot} \right)^{-1} \yr,
\label{tkh2}
\end{equation}
where $M_{\rm env}$ is the envelope mass.

For the relevant dynamical time I take $(G \rho_{\rm av})^{-1/2}$,
where $\rho_{\rm av}$ is the average density of the entire star.
Scaling with typical number gives
\begin{equation}
\tau_{\rm dyn} \simeq 0.7
\left( \frac{M}{0.6 M_\odot} \right)^{-1/2}
\left( \frac{R}{300 R_\odot} \right)^{3/2} \yr.
\label{tdyn1}
\end{equation}

During the evolution along the AGB before the star starts to contract in radius, the luminosity increases
and the mass decreases, such that $\tau_{\rm dyn}$ increases and $\tau_{\rm KH-env}$ decreases.
Therefore, their ratio $Q \equiv \tau_{\rm dyn}/\tau_{\rm KH-env}$
increases.
For evaluating the value of $Q$ on the upper AGB,
when the envelope mass is low, I take $M_c=M$ in equation (\ref{tkh2}).
This gives for the upper AGB
\begin{equation}
Q \equiv \frac{\tau_{\rm dyn}}{\tau_{\rm KH-env}}
\simeq 0.1
\left( \frac{\beta_s}{5} \right)^{-1}
\left( \frac{M}{0.6 M_\odot} \right)^{-3/2}
\left( \frac{R}{300 R_\odot} \right)^{5/2}
\left( \frac{L}{5000 L_\odot} \right)
\left( \frac{M_{\rm env}}{0.1 M_\odot} \right)^{-1}
\label{chi1}
\end{equation}
The thermal time $\tau_{\rm KH-env}$ is less than an order of magnitude longer
than the dynamical time during this late AGB phase.
Soker \& Harpaz (1999) noted that this relatively short thermal time must
result in a strong irregular behavior of the envelope because dynamical motions,
such as pulsations and convective motion, can cause large thermal perturbations in the envelope.
Therefore, a large value of this parameter, i.e., $Q \ga 0.1$, can be related to
a highly inhomogeneous mass loss process, as well as to a high mass loss rate.
Soker \& Harpaz (1992) argued that the characteristics of AGB stellar pulsations
depend on the thermal and dynamical time scales. They used the thermal time scale of only
the upper envelope, and for the dynamical time scale they took the pulsation period.

The star starts to contract before it leaves the AGB. The initial contraction is slow,
and the stellar radius during the early contraction phase can be
approximated by
\begin{equation}
R_{C} \simeq R_m
\left( \frac{M_{\rm env}}{M_{\rm env-m}} \right)^{\delta},
\label{rad1}
\end{equation}
where $R_m$ and $M_{\rm env-m}$ are the stellar radius and envelope mass when the
contraction starts.
For the model used by Soker (1992) relation (\ref{rad1}) holds for an envelope
mass of $0.001 \la M_{\rm env} \la 0.1 M_\odot$ with $\delta \simeq 0.2$ for most of the time.
Then $\delta$ increases more and more rapidly until it reaches a very large value
when the star contracts by two order of magnitude for a tiny change in the envelope mass
(Sch\"onberner 1983).
Qualitatively similar behavior is found for other core masses, but at different envelope masses
(Sch\"onberner 1983; Frankowski 2003).
Using equation (\ref{rad1}) to express the stellar radius in equation (\ref{chi1})
gives for the contracting-AGB phase
\begin{equation}
Q_{ C}
\simeq 0.1
\left( \frac{\beta_s}{5} \right)^{-1}
\left( \frac{M}{0.6 M_\odot} \right)^{-3/2}
\left( \frac{L}{5000 L_\odot} \right)
\left( \frac{M_{\rm env}}{0.1 M_\odot} \right)^{\frac{5}{2}\delta-1}
\left( \frac{R_m}{300 R_\odot} \right)^{5/2}
\left( \frac{M_{\rm env-m}}{0.1 M_\odot} \right)^{-\frac{5}{2}\delta}.
\label{chi2}
\end{equation}

During the contracting-AGB phase the luminosity and mass do not change much,
and the derivative of equation (\ref{chi2}) can be written as
\begin{equation}
\Delta Q_C \equiv \frac{d \ln Q_{C}}{d \ln M_{\rm env}} \simeq
\frac{5}{2}\delta-1
- \frac{d \ln \beta_s}{d \ln M_{\rm env}}.
\label{chi3}
\end{equation}
Along the entire AGB $\Delta Q$ is negative (beside temporal variations, e.g.,
after thermal pulses). It is very positive during the fast contraction along the
post-AGB track (again, beside temporal variations).
The transition from $\Delta Q <0$ to $\Delta Q>0$ can mark the beginning of
the post-AGB phase.
(Note that the envelope mass decreases with time, and therefore when $\Delta Q <0$
then $Q$ increases with time.)
Namely, the star is said to terminate the AGB when $Q$ is at its maximum value; this
occurs after the contraction started and $Q=Q_C$.

If there is no change in the density profile then $d \ln \beta_s/d \ln M_{\rm env} = 0$
and $\Delta Q_C$ changes sign when $\delta = 0.4$.
This is when more or less the rapid contraction starts.
However, during the contracting-AGB phase the density profile becomes steeper
(e.g., Soker 1992), and $\beta_s$ increases slowly, so that
$d \ln \beta_s/d \ln M_{\rm env} < 0$.
On the other hand, the envelope convective zone, which might be more relevant to
many processes influencing the mass loss process, becomes concentrated in the outer region
and the effective value of $\beta_s$ might decrease.
Over all, I suggest to ignore the structural factor $\beta_s$, and
to mark the transition when $\delta = 0.4$, i.e., when
$d \ln (R_{C})/ d \ln {M_{\rm env}} = 0.4$ and $\Delta Q=0$.

Alternatively we can use the criterion that
\begin{equation}
\Delta Q_{\rm KH} \equiv - \frac{d \ln \tau_{\rm KH-env}}{d \ln M_{\rm env}} \simeq
\delta-1
- \frac{d \ln \beta_s}{d \ln M_{\rm env}}
\label{chi4}
\end{equation}
is equal zero. Namely, $\tau_{\rm KH-env}$ has its minimum value.
This occurs when $\delta \simeq 1$.

In the model presented in Soker (1992)   
$\delta = 2/5$ and $\delta = 1$ when the
stellar radius is $R=150 R_\odot$ and $R=110 R_\odot$, and the effective temperature is
$T=4400\K$ and $T=5100\K$, respectively.
The value of $C$ defined in equation (\ref{C1})  reaches its maximum at $T \simeq 7000 \K$.
In a model with a core mass of $M_c=0.67 M_\odot$ from the solar metallicity track
of Vassiliadis \& Wood (1994), the maxima in $Q$ and $Q_{\rm KH}$ as calculated
by Frankowski (Frankowski, A. 2007, private communication) are reached
when the stellar radius is $R=120 R_\odot$ and $R=85 R_\odot$,
and at an effective temperature of $T=5200\K$ and $T = 6100 \K$, respectively.
The value of $C$ has its maximum at a temperature of $T = 8900 \K$ in that model.
As mentioned in section 1, the $C$-criterion gives the transition at a too high
temperature, after mass loss has decline, but still captures most aspects of the transition.
I prefer the $Q$-criterion or the $Q_{\rm KH}$-criterion because they seem to have more
of a physical implication to the mass loss process.

\section{SUMMARY}
\label{sec:summary}

I suggest to theoretically define the transition from the AGB to the post-AGB phase,
namely, the termination of the AGB, by using the envelope thermal time scale
(Kelvin-Helmholtz time) and the stellar dynamical time.
The criterion does not refer to the short time-scale variations occurring on the
AGB and post-AGB, e.g., thermal pulses (helium shell flashes) and magnetic activity,
but refers to the time average of stellar properties.
All other alternatives for the transition must use average values as well;
only a criterion based on the envelope mass alone does not need a time average,
but no such a criterion has been suggested.

As the star expands along the AGB and loses mass the thermal time scale decreases and the
dynamical time increase. On the upper AGB the two time scales become comparable.
This implies that dynamical processes, such as pulsation and convective motion, can influence
the thermal state of the envelope. This is likely to influence the mass loss process.
The thermal time scale continues to decrease even after the star starts its contraction.
The thermal time scale starts to increase at about the same evolutionary point
where the rapid contraction starts.

I suggest to define the termination of the AGB when $Q$ (defined in eq. \ref{chi1})
reaches it maximum value.
This occurs during the contraction part of the AGB, when $Q=Q_C$ (eq. \ref{chi2}),
and the transition occurs when $\Delta Q_C=0$ (defined in eq. \ref{chi3}).
Alternatively, using only the thermal time scale, the transition can be defined when
the thermal time scale starts to increase with decreasing envelope mass.
Namely, when the value of $\tau_{\rm KH-env}$ is at its minimum and
$\Delta Q_{\rm KH}=0$ (eq. \ref{chi4}).

The criterion proposed here has several advantages.
\begin{enumerate}
\item It uses well defined properties of the envelope. These
are simple to derive with a stellar model calculated without the inclusion
of pulsation or magnetic activity. If these are included, time average must
be introduced.
\item The transition can be defined to occur at a well define
 evolutionary point (when averaged over short time scales variations), when
 $\Delta Q=0$ (or, alternatively, $\Delta Q_{\rm KH} =0 $).
\item It is based on properties that are closely related to the mass loss process:
 The dynamical time and the thermal properties of the envelope.
However, the theoretical transition occurs when $Q$ is at its maximum value. Therefore,
the high mass loss rate and inhomogeneities are likely to continue into the post-AGB
phase for a short time.
\item It contains previously proposed criteria.
 (a) Pulsation: via the dynamical time scale. (b) Rapid contraction:
 as the change in the behavior of ${\tau_{\rm KH-env}}$ occurs
at about the same time rapid contraction stars.
(c) Mass loss rate: The mass loss process is related to the dynamical and thermal time scales.
(d) Effective temperature: The value $\Delta Q =0$ is reached at $T \sim 4000-6000 \K$,
a temperatures range that was used before to mark the zero age pos-AGB phase.
\end{enumerate}

Although the criterion proposed here has its own merit in the theoretical study of
stellar evolution, it goes beyond a pure academic exercise.
The criterion suggests that the involved time scales have a role in determining
the evolution of the envelope, mainly as they are important parameters in the mass loss
process, e.g., via pulsation properties.
Future theoretical studies will have to examine in more detail how these time scales affect the mass
loss process.
Also, the suggested theoretical criterion must be examined with detailed numerical
simulations, to verify that the criterion $\Delta Q=0$, or $\Delta Q_{\rm KH} =0$,
do not give wrong results.
The numerical study should also determine which of these two similar criteria
fit observational parameters better, hence connecting theoretical studies with
observations.

\acknowledgements
This research was supported by the Asher Space Research Institute in the Technion.

\end{document}